\documentclass[lettersize,journal]{IEEEtran}
\usepackage{amsmath,amsfonts}
\usepackage{algorithmic}
\usepackage{algorithm}
\usepackage{array}
\usepackage[caption=false,font=normalsize,labelfont=sf,textfont=sf]{subfig}
\usepackage{textcomp}
\usepackage{stfloats}
\usepackage{url}
\usepackage{verbatim}
\usepackage{graphicx}
\usepackage{cite}
\newcolumntype{H}{>{\setbox0=\hbox\bgroup}c<{\egroup}@{}}

\usepackage{xurl}
\usepackage{booktabs}
\usepackage{xcolor}
\usepackage{csquotes}
\usepackage{multirow}
\usepackage{enumitem}

\newcommand{\eg}{\textit{e.g.}}
\newcommand{\ie}{\textit{i.e.}}
\newcommand{\etal}{\textit{et~al.}}

\newcommand{\revise}[1]{\textcolor{black}{#1}}
\newcommand{\re}[1]{\textcolor{black}{#1}}
\newcommand{\rem}[1]{\textcolor{black}{#1}}

\begin{document}

\title{Why is AI not a Panacea for Data Workers? An Interview Study on Human-AI Collaboration in Data Storytelling}

\author{%
  Haotian Li, Yun Wang, Q. Vera Liao, and Huamin Qu
  
\thanks{

H. Li and H. Qu are with the Hong Kong University of Science and Technology. E-mail: haotian.li@connect.ust.hk and huamin@cse.ust.hk.

Y. Wang is with Microsoft Research Asia. Y. Wang is the corresponding author. E-mail: wangyun@microsoft.com. 

Q. V. Liao is with Microsoft Research Montréal. E-mail: veraliao@microsoft.com.

\rem{This paper is part of the thesis ``Bridging Data Analysis And Storytelling With Human-AI Collaborative Tools'' by the first author H. Li~\cite{li2024bridge}.}
}
  
}

\markboth{Journal of \LaTeX\ Class Files,~Vol.~14, No.~8, August~2021}%
{Shell \MakeLowercase{\textit{et al.}}: A Sample Article Using IEEEtran.cls for IEEE Journals}


\maketitle

\begin{abstract}
This paper explores the potential for human-AI collaboration in the context of data storytelling for data workers.
Data storytelling communicates insights and knowledge from data analysis. It plays a vital role in data workers' daily jobs since it boosts team collaboration and public communication.
However, to make an appealing data story, data workers need to spend tremendous effort on various tasks, including outlining and styling the story.
Recently, a growing research trend has been exploring how to assist data storytelling with advanced artificial intelligence~(AI).
However, existing studies focus more on individual tasks in the workflow of data storytelling and do not reveal a complete picture of humans' preference for collaborating with AI.
To address this gap, we conducted an interview study with 18 data workers to explore their preferences for AI collaboration in the planning, implementation, and communication stages of their workflow. 
We propose a framework for expected AI collaborators' roles, categorize people’s expectations for the level of automation for different tasks, and delve into the reasons behind them. 
Our research provides insights and suggestions for the design of future AI-powered data storytelling tools.
\end{abstract}

\begin{IEEEkeywords}
Data storytelling, Human-AI collaboration, Interview study
\end{IEEEkeywords}





\section{Introduction}
\re{Data workers, such as business analysts, data journalists, and data scientists, perform data analysis and exchange results from analysis as two essential parts of their daily work. 
To communicate insights and knowledge effectively and engagingly, data storytelling, \rem{\ie, creating and communicating a series of coherently connected data facts with visualizations and narratives to achieve the author's goal~\cite{lee2015more}}, has been leveraged to boost collaboration in teams~\cite{brehmer2022jam}, 
raise public awareness~\cite{lee2015more}, and so on.}

However, \re{telling data stories} is non-trivial for data workers.
First, data workers need to master various skills and tools for authoring data stories, such as authoring charts and writing narrations~\cite{lee2015more}.
Furthermore, the knowledge gap between data workers and their diverse audiences 
even makes the task more challenging~\cite{piorkowski2021ai,mao2019data}.
Data workers have to spend substantial effort preparing clear, coherent, and engaging data stories to effectively communicate their identified data insights and knowledge to the target audience. 
Informed by the advance of artificial intelligence~(AI), including the recent development of large-scale AI systems (\eg, DALL·E~\cite{dalle} and ChatGPT~\cite{chatgpt}),
researchers  
have attempted to tackle the challenges by introducing AI-powered data storytelling tools. 
\re{Nevertheless, most of the existing research studies are limited to studying human-AI collaboration~\cite{wang2020human} and developing intelligent tools for specific data story components (\eg, color~\cite{yuan2021infocolorizer, shi2022colorcook} and animation~\cite{ge2021cast, kim2021gemini}) or scenarios (\eg, computational notebooks~\cite{zheng2022telling, li2023notable} and visual analytics systems~\cite{chen2018supporting, gratzl2016visual}).}
These studies \rem{lack} an overall picture of 
where and how
humans would like to collaborate with AI in the entire workflow.

To better leverage the powerful AI capabilities in intelligent tools, the philosophy of human-centered design \cite{cooley2000human} informs the necessity of a comprehensive understanding of data workers' needs, workflows, and attitudes toward AI, rather than making assumptions about where, how, and even whether these tools should be developed. 
With this in mind, we conducted an interview study involving 18 data workers---people who work on data, including researchers, data scientists, and business analysts, aiming to illuminate future research directions for intelligent tools, particularly in the context of supporting data storytelling.
\re{Our study is guided by three research questions regarding the data storytelling workflow:}
\begin{enumerate}
    \item \textit{Where would data workers like to collaborate with AI?}
    \item \textit{How would data workers like to collaborate with AI?}
    \item \textit{Why and why not do data workers prefer to collaborate with AI?}
\end{enumerate}
\rem{By answering the first and second questions, we hope to identify and differentiate expected collaboration approaches with AI in various tasks of the entire workflow.
The rationales behind these preferences can be revealed through the answers to the third question.
Combining these insights together, we expect to provide suggestions for future tool design, including the desired features and the pitfalls to be avoided.}



Our findings provide answers to these questions as follows: 
we first summarize participants' workflow of data storytelling, which involves three stages, \textit{planning}, \textit{implementation}, and \textit{communication}, and seven tasks. 
Based on this workflow, we outline four different AI collaborator roles described by participants and participants' preferences around these roles.
Specifically, these roles are \textit{creator}, \textit{optimizer}, \textit{reviewer}, and \textit{assistant}, in the order of \re{decreasing automation levels of AI}. 
Furthermore, we report the participants’ reasons why or why not AI is desired in their data storytelling workflows. For example, AI was appreciated due to its quick turnover time but criticized for its limited ability to understand humans’ contexts. To conclude our research, we discuss the implications for future human-AI collaborative data storytelling tool design.


\section{Related work}
\rem{Our research is built on the empirical research on human-AI collaborative data science, AI-powered data storytelling tools, and data storytelling-related empirical studies.}


\subsection{Empirical research on human-AI collaborative data science}\label{sec:related_data_science}

\re{Currently, data science tasks are still labor intensive and require deep domain expertise. 
A typical data science workflow includes preparing data, exploring data, modeling data, communicating data insights, and deploying models~\cite{zhang2020data,muller2019data, kandel2012enterprise, alspaugh2018futzing, crisan2020passing}.} 
\rem{To assist these tasks in the workflow, there is growing interest in introducing AI to support humans' work, such as automatic data preparation and modeling (\ie, AutoML)~\cite{narkar2021model,weidele2020autoaiviz}, AI-assisted data exploration ~\cite{setlur2016eviza,wu2021ai4vis}, and AI-powered data storytelling for insight communication (Sec.~\ref{sec:ai_storytelling_tools}).}


\rem{To design AI collaborators that fit humans' needs, it is essential for HCI researchers to investigate data workers' perceptions, requirements, and concerns for AI through empirical studies.
For example, Gu~\etal~\cite{gu2024analysts} developed design probes and investigated data workers' behaviors when exploring data with AI.}
Wang~\etal~\cite{wang2019human} conducted an interview study with 20 data scientists about their perception of AutoML tools and discovered their concerns around job security and desire for human-AI collaboration, in which both automation and human expertise of data science are indispensable.
Focusing on data workers' trust in AutoML systems, Drozdal~\etal's~\cite{drozdal2020trust} study revealed a strong need for transparency features such as performance metrics and visualization to establish trust in AutoML.
By interviewing 29 enterprise data scientists, Crisan and Fiore-Gartland~\cite{crisan2021fits} identified common usage scenarios of AutoML systems and provided a framework summarizing the level of automation desired by data workers. 

\rem{In addition to data preparation, exploration, and modeling, communicating insights with data storytelling plays an important role for data team cooperation~\cite{brehmer2022jam, piorkowski2021ai, wang2019data} and communication with clients~\cite{hullman2013deeper}.
Telling data stories requires data workers to author story pieces with visualizations and narratives to introduce their identified data-related facts, then connect the story pieces following a coherent narrative structure, and finally communicate the organized story pieces with targeted audiences~\cite{lee2015more}. 
Though various AI-powered tools for data storytelling exist, we found that there lacks empirical studies to investigate where, how, and why (not) data workers would like to collaborate with AI in data storytelling.
We envision that a systematic investigation on these topics can surface data workers' requirements and provide design guidelines for future tools in a holistic manner.
}



\subsection{AI-powered data storytelling tools}\label{sec:ai_storytelling_tools}
Data storytelling and narrative visualization tools have become increasingly popular in recent years. These tools aim to help users create compelling data stories by providing data-driven insights and suggestions for visualization. AI in these tools has the potential to significantly enhance the storytelling process by automating tasks in the data storytelling workflow.

\re{To enhance the workflow of data storytelling, researchers have used AI techniques in various approaches. 
They have applied vision techniques to understand the visual structure and semantics of infographics \cite{lu2020exploring, wang2021animated}.
AI techniques are also used for different dimensions of design to make data storytelling more effective and engaging for communicating the intended message. 
The dimensions include color \cite{yuan2021infocolorizer, shi2022colorcook}, layout \cite{qian2020retrieve}, graphics \cite{kim2016data, wang2018infonice}, motions and animations \cite{wang2021animated, ge2021cast}.
Researchers also try to 
automate or semi-automate the process of creating and recommending data storytelling in the form of infographics \cite{cui2019text}, 
data videos \cite{shi2021autoclips, shin2022roslingifier}, data comics \cite{kang2021toonnote}, data articles~\cite{sultanum2023datatales}, etc., under different scenarios, including computational notebooks~\cite{zheng2022telling, li2023notable} and visual analytics systems~\cite{chen2018supporting, gratzl2016visual}. 
For better results, existing research further attempts to evaluate visual designs in data stories with AI~\cite{fu2019visualization}.}

\re{These studies have contributed various experiences of introducing AI into the data storytelling workflow.
Considering their research scope, they either serve for authoring a specific component of data stories, such as color~\cite{yuan2021infocolorizer, shi2022colorcook}, or a certain usage scenario, such as data analysis with computational notebooks~\cite{lin2023inksight, li2023notable}.}
As a result, they do not provide a holistic view of data workers' expectations and challenges for collaborating with AI, such as why and why not humans would like to collaborate with AI.
In our interview, we focus on users' expectations for human-AI collaboration without the constraints of specific data story components or scenarios and provide insights about introducing AI to future tools.
\rem{Through a more comprehensive investigation, we can identify various approaches to collaborate with AI and address the diverse needs within a single task. Additionally, general guidelines for applying AI models can be derived to benefit tool design.}

\subsection{Empirical research on data storytelling}\label{sec:storytelling_empirical}
\re{Empirical research serves as an important approach for distilling knowledge from existing data storytelling practices and benefiting future data story authoring.
Such research includes design spaces of data stories (\eg,~\cite{yang2021design, hao2024design}), perceptions of data stories~(\eg,~\cite{borkin2013makes, shu2020makes}), and practices of data storytelling~(\eg,~\cite{chevalier2018analysis, brehmer2022jam}).
Our research is closest to the empirical research on data storytelling practices.}

\re{As a pioneer study, Chevalier~\etal~\cite{chevalier2018analysis} understood the workflow of data storytelling with an interview with nine practitioners.
They summarized three stages of data storytelling, including exploring data, preparing stories, and telling stories.
Showkat and Baumer~\cite{showkat2021stories} focused on the practices of data exploration in data storytelling workflow proposed by Chevalier~\etal~\cite{chevalier2018analysis} and Lee~\etal~\cite{lee2015more}.
A recent work by Brehmer and Kosara~\cite{brehmer2022jam} examined the practices of communicating data in organizations and identified three scenarios from informal team meetings to formal presentations.
Other visualization design practice research also reflects data storytelling practices partially, such as activities in visualization design~\cite{mckenna2014design}.}

The previous research focuses on understanding the existing workflow of data storytelling with different emphases.
These studies do not surface data storytelling practitioners' expectations and concerns for AI assistance.
\rem{With more knowledge about their attitudes towards AI, it is possible to design tailored AI-powered tools that address their challenges effectively and efficiently.}
To achieve the goal, grounded in the data storytelling workflow summarized from our interviews, we describe a series of insights regarding where, how, and why (not) data workers would like to collaborate with AI.

\section{Methodology}
Our study was conducted in the first quarter of 2023 when multiple breakthrough AI systems appeared, such as ChatGPT~\cite{chatgpt}, new Bing~\cite{newbing}, and GPT-4~\cite{openai2023gpt4}.
These systems might lead to data workers' changing attitudes toward AI and encouraged us to investigate their needs and ideas about human-AI collaboration.

\subsection{Participant recruitment and background}\label{sec:participants}
In our study, we recruited data workers from both academia and industry. 
\revise{Data workers have various definitions in previous studies.
For example, Boukhelifa~\etal~\cite{boukhelifa2017data} and Liu~\etal~\cite{liu2019understanding} define data workers as those who are amateur data analysts and are not self-identified as data scientists.
Two recent studies~\cite{zhang2020data, muller2019data} noticed that the boundary between professional data scientists and amateur data analysts is unclear and named all people who do data science work as ``data science workers''.
In our research, we follow the second definition of data workers to include  people who work on data, including researchers, data scientists, and business analysts.
The definition can reduce the need of drawing a rigorous boundary between data science professionals and amateurs and make our study more inclusive.}

To recruit data workers for our interviews, we adopted multiple approaches, including posting ads on social media or in special interest groups through communication software and sending invitations leveraging our professional network.
\re{We ensured that the participants had
experiences in data storytelling by directly asking whether they had created data stories in their daily work or confirming that telling data stories is part of their job based on their job natures before sending out invitations.
Their detailed experience, \eg, years in data work and data storytelling workflow, was further enquired in interviews through self-reporting to \rem{check participants' eligibility for the study.}}

Ultimately, we recruited 18 participants (12 males and 6 females, 9 from academia and 9 from the industry, average experience in data analysis = 5.39, the standard deviation of experience in data analysis = 3.10).
P6 and P17 \rem{were} postgraduate researchers, but they mainly talked about their data storytelling-related experiences in the interviews when they took full-time duties in the industry. 
Therefore, we only reported their industrial backgrounds.
\re{The most common domain of our participants is data science (six participants), followed by finance (three participants) and consulting (two participants). 
Most of them frequently use slide decks as the format of their data stories.
Some other formats of data stories include reports and articles.
When authoring data stories, they used slide tools (all participants), including Microsoft Powerpoint and Google Slides, and programming tools, such as Python and D3 (ten participants).
Microsoft Excel is also a common tool used by six participants.
Table~\ref{tab:demographic} shows detailed demographic information.}

\revise{Besides experience in data storytelling, all participants have AI-related experiences. Four participants have experience in applying AI for data storytelling and other creative work. P1, P8, and P17 have used the design idea~\cite{designidea} function of Powerpoint.
P3 has often applied generative models to create artwork. 
Furthermore, seven participants (P8, P9, P12, P14, P15, P16, and P17) have used chatbots like ChatGPT and new Bing. 
For example, P16 sought ChatGPT's help when writing emails. 
P1, P7, P8, P9, and P18 have conducted research on AI-powered tools.
P2, P4, P6, P13, and P15 have developed or researched AI models.
The experience of P5, P10, and P11 is limited to AI-related courses. }



\begin{table*}[h!]
    \centering
    \caption{ 
    \revise{The table records our interviewees' demographic information, including genders, ranges of ages, jobs, domains, and experiences of data work in years (denoted as Exp. in the table), frequently used genres of data stories (Genre), and tools for making data stories (Tools). When reporting tools, since Google Slides and Powerpoint are both frequently mentioned for authoring slides, we unify them as \textit{``slide tools''} for brevity. Similarly, we use \textit{``programming tools''} to unify Python, Vega, D3, etc.}}
    \scriptsize
    \begin{tabular}{@{}lllllllp{6.5cm}HH@{}}
    
    \toprule
      \textbf{ID} & \textbf{Gender} & \textbf{Age} & \textbf{Job} & \textbf{Domain} & \textbf{Exp.} & \textbf{Genre} & \textbf{Tools} & \textbf{AI Experience} & Name\\ \midrule
      1 & Male & 25-30 & Researcher & Data Science & 6 & Slides & Slide Tools & Design Ideas & Aoyu \\
      2 & Male & 40-45 & Software Engineer & Data Science & 10 & Slides & Slide Tools, Excel, Paper & Design Ideas & Borje \\
      3 & Male & 25-30 & Researcher & Chemistry & 4 & Slides & \re{Slide Tools, Excel, Programming Tools, OriginLab, ImageJ, Igor} & Design Ideas & Jiakai \\
      4 &  Male & 25-30 & Researcher & Data Science & 6 & Slides & Slide Tools, Programming Tools, Paper & Design Ideas & Jiaqi \\
      5 & Female & 25-30 & Researcher & Algorithm & 3 & Slides & Slide Tools, Programming Tools & Design Ideas & Juanru \\
      6 & Male & 25-30 & Data Scientist & Consulting & 4.5 & Slides, Interactive Demo & Slide Tools, Excel & Design Ideas & Kentaro \\
      7 & Male & 25-30 & Researcher & Data Science & 6 & Slides, Video & Slide Tools, Programming Tools, iMovie & Design Ideas & Leixian \\
      8 & Female & 25-30 & Researcher & Data Science & 5 & Slides & Slide Tools, Tableau, Programming Tools, Paper & Design Ideas & Leni \\
      9 & Female & 25-30 & Researcher & Data Science & 3 & Slides, Video, Infographics & Slide Tools, Figma, Illustrator, Programming Tools, iMovie & Design Ideas & Lu \\
      10 & Female & 25-30 & Data Analyst & Finance & 4 & Slides & Slide Tools, Programming Tools& Design Ideas & Min \\
      11 & Male & 25-30 & Business Analyst & Finance & 1.5 & Slides & Slide Tools & Design Ideas & Mingyuan \\
      12 & Male & 25-30 & Business Analyst & Consulting & 7 & Slides & Slide Tools & Design Ideas &  Peiyang \\
      13 & Male & 25-30 & Researcher & HCI & 3 & Slides & Slide Tools, Word, Excel, Programming Tools & Design Ideas & Qingyu \\
      14 & Male & 25-30 & Student & Economics & 5 & Slides, Reports & Slide Tools, Word, Excel, Programming Tools & Design Ideas & Rongtao \\
      15 & Male & 25-30 & Researcher & Medical Imaging & 3 & Slides & Slide Tools, Programming Tools & Design Ideas & Tianyuan \\
      16 & Female & 30-35 & Business Analyst & Finance & 8 & Slides & Slide Tools, Excel & Design Ideas & Ye \\
      17 & Female & 25-30 & Journalist & Journalism & 3 & Articles & Slide Tools, PhotoShop, Word, Platform-provided Text Editors & Design Ideas & Yifan \\
      18 & Male & 40-45 & Applied Researcher & Software & 15 & Slides, Videos & Slide Tools, Expression Studio & Design Ideas &  Zhitao \\  
    \bottomrule
    \end{tabular}
    \label{tab:demographic}
\end{table*}

\subsection{Interview procedure}\label{sec:interview}
In our study, we conducted semi-structured interviews with the participants through online meeting software, such as Zoom and Microsoft Teams, per participants' convenience.
Each interview involved one or two authors and the participant.
\rem{The interviews were conducted in either English or Chinese, based on participants' preferences.}

Before the interview, we first introduce the purpose of our study.
Then we collected participants' consent to participation in the study and recorded the rest of the meeting.
All interviews started with a query about the participants' data storytelling-related background, such as their job nature.
\re{Then, we moved to our main questions relating to their data storytelling practices and desired assistance from AI in creating data stories.}
\rem{To present all interview materials clearly and organized, we listed them in a slide deck available in our supplementary material. This slide deck was shown to our participants in all three parts of our interview.}

In the first part, we asked the participants about their familiar format of data story and then invited them to share a recent piece of their stories with us if convenient.
Then we created mind maps with the participants to help them recall and organize their workflow of data storytelling.
\re{Specifically, we provided a skeleton of the mind map using Xmind\footnote{\url{https://xmind.app/}} to indicate the information we would like to collect, \ie, stages in their workflow and tasks and tools in each stage.
Then the mind map is created either by the participants directly or by the authors following the participants' description.
The creation of mind maps was shared between interviewees and authors through screen sharing to make sure that (1) the mind map followed the participants' opinions accurately when it was created by the authors and (2) the authors understood the mind map clearly when it was created by interviewees.
During the mind map creation, the authors may also ask the participants about more details of their workflow.}

The second part began with examples of AI applications in data storytelling to help brainstorm.
\re{Following the approach of a previous interview about users' attitudes towards human-AI collaboration in \rem{another scenario~\cite{jiang2021supporting}}, we encouraged them to consider beyond the existing advanced AI models (\eg, ChatGPT).
\re{Such a design also aligns with the idea of speculative design~\cite{dunne2024speculative}, provoking the discussion of future challenges and solutions to better prepare us for the fast-growing large-scale AI systems.}
In this way, we hoped to provide useful and long-lasting references for future tool design without the restriction of the currently available AI techniques.}
Then the participants freely proposed where and how they prefer to collaborate with AI for data storytelling.
\re{The ideas were also documented in the mind map to help them organize their thoughts.}
We further probed on the reasons behind their preferences for involving AI in different ways.

\rem{Once the participants were satisfied with their ideas, we moved to the last part of the interview.}
In this part, we asked about overall perceptions of human-AI collaboration in data storytelling, such as the difference between collaboration with AI and human designers. 
Each interview lasted for around 0.5-1 hour.
All interviews were recorded and transcribed. 
The average length of recorded videos is about \re{43 minutes.}

\subsection{Result analysis}\label{sec:analysis}
We adopted an iterative coding approach when analyzing our participants' responses.
At the beginning of our study, two leading authors attended the first six interviews together to ensure a common understanding of the protocol and its coverage of the research questions.
\revise{Following previous research methodology~\cite{moore2023failurenotes, mcnutt2023design}, the first author first open-coded the transcripts for the desired collaboration patterns (\ie, where and how the participants would like to collaborate with AI) and reasons behind these patterns after the interviews.
Then the codes were reviewed by the second author and discussed between the two authors iteratively until reaching an agreement.}
Along with more interviews conducted, the collaboration patterns and reasons were examined, refined, merged, and enriched.
In the next section, we present the interview results supported by participant quotes.

\section{Results}
This section targets answering the research questions with interview results. 
\rem{Sec.~\ref{sec:workflow} introduces the participants' workflow to ground results for our research questions.
Then, based on why or why not~(Secs.~\ref{sec:why} and~\ref{sec:why_not}) data workers would like to collaborate with AI, we introduce where and how (Sec.~\ref{sec:where_and_how}) they expect the collaboration should be.}


\begin{figure}[h!]
    \centering
    \includegraphics[width=\linewidth]{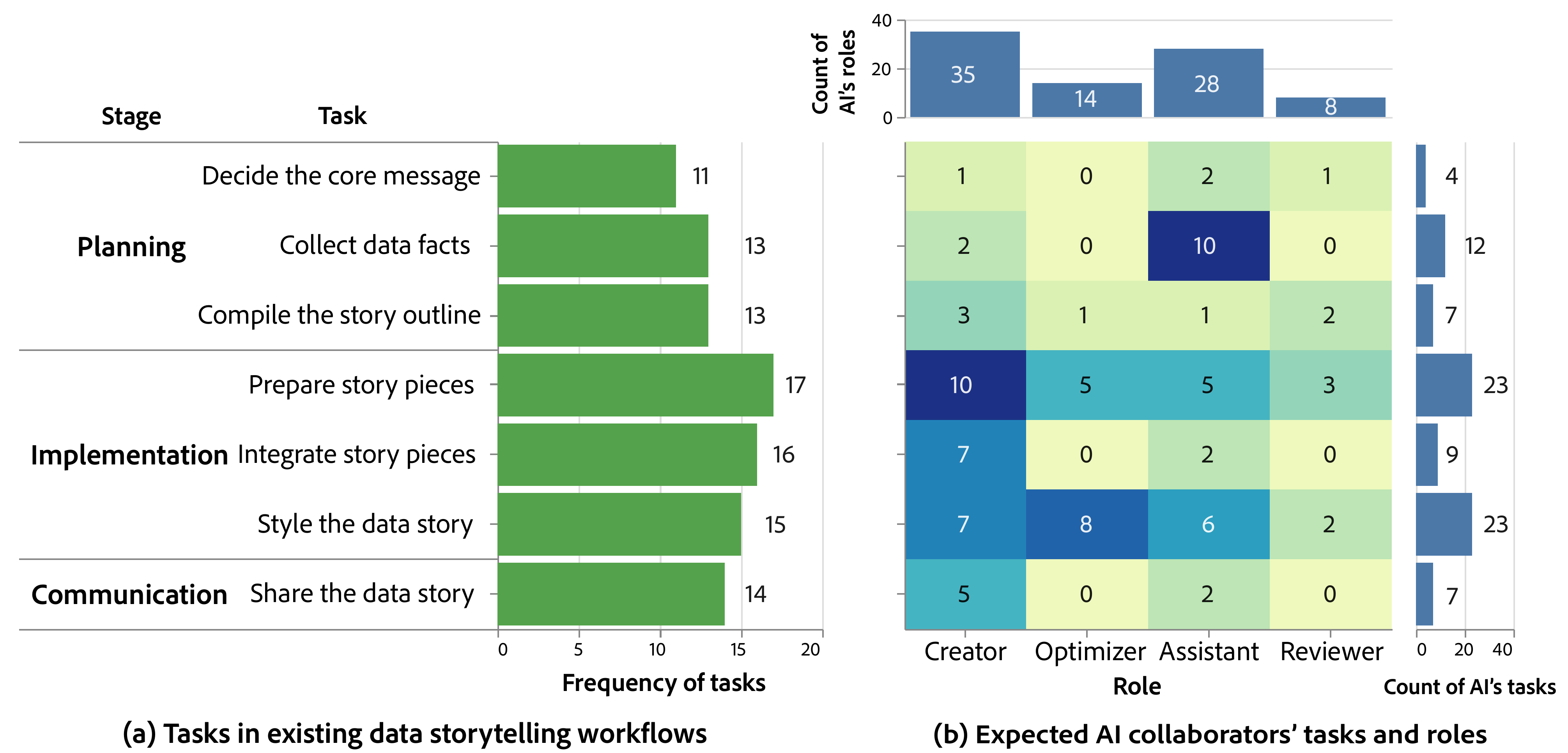}
    \caption{
    This figure summarizes participants' opinions about where and how they would like to collaborate with AI.
    (a) shows tasks in the existing workflows of our interviewees and (b) illustrates the expected AI collaborators' roles and their tasks.
    Both y-axes in (a) and (b) are the tasks in the data storytelling workflow. In (b), the heatmap presents the breakdown frequency of task-role tuples. The bar chart on the top counts the frequency of roles, while the one on the right shows the frequency of tasks. Notably, since each participant can propose multiple roles of AI collaborators for one task, the count of tasks can be larger than the number of participants. 
    }
    \vspace{-1em}
    \label{fig:task-role}
\end{figure}

\begin{figure*}
    \centering
    \includegraphics[width=\linewidth]{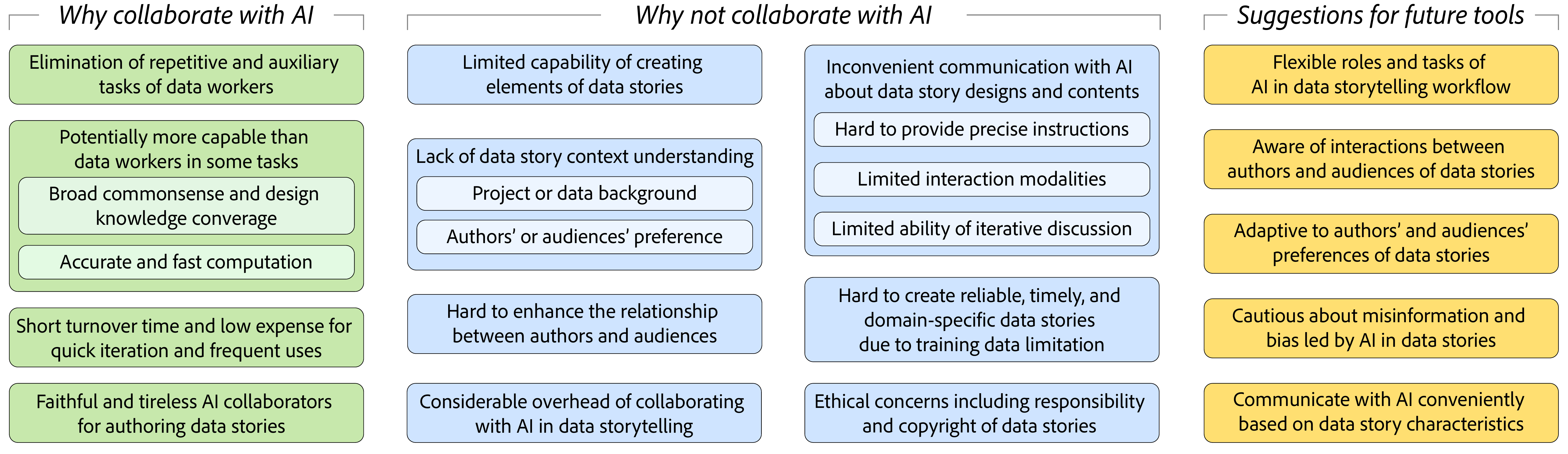}
    \caption{This figure summarizes (1) the interviewees' opinions about why and why not AI collaborators are preferred and (2) our suggestions for future AI-powered data storytelling tools.
    }
    \vspace{-1em}
    \label{fig:why_why_not}
\end{figure*}

\subsection{What is the general workflow of data storytelling?}\label{sec:workflow}
\re{We considered that there could be various dimensions to summarize the workflow, such as the purpose of each task from a procedural perspective and the scenarios of storytelling (\eg, interpersonal or mass communication).
As Sec.~\ref{sec:ai_storytelling_tools} states, we would like to release the constraints of scenarios and provide an overview of data workers' expectations for AI.
Therefore, we characterized the workflow with task purposes.}
\re{According to our participants, their workflow involves three stages and seven tasks, aligning with prior research~\cite{lee2015more, chevalier2018analysis}.}
\re{They often first \textit{plan} how to tell a data story, then \textit{implement} their plan to instantiate the outline to a complete story, and finally \textit{communicate} the data story with others.}
\re{Notably, the stages and tasks are often conducted iteratively rather than linearly, according to our participants.
Fig.~\ref{fig:task-role}(a) presents the task frequency mentioned by our interviewees.}

\re{The first stage in the workflow is \textbf{planning}, where the interviewees comprehend their results from data analysis and brainstorm the overall picture of a data story.
The interviewees commonly first \textit{decide on the core message of the data story} based on their data findings and the broader context of their projects, such as the background and the target.
\rem{Next, they move on to \textit{collecting supportive data facts} that are directly used in data stories.
Data facts can be derived from data analysis by data workers themselves or collected from other resources, such as reports.}
\rem{To support or interpret these data facts, it is sometimes necessary to gather additional information from other materials, \eg, news.}
The last task in the stage is to \textit{compile the story outline}, where the related information is organized based on topics and logical relationships.}

\re{In the second stage, the interviewees \textbf{implement} the plans created in the first stage.
They commonly first \textit{prepare story pieces} to transform collected facts into a presentable format, such as drafting the introduction to data analysis methods and making charts.
The next task is to \textit{integrate story pieces},
where the interviewees make presentation slides or reports to accommodate the story pieces according to the outline.
\rem{At the end of the implementation stage, the interviewees often \textit{style the data story} from various aspects.
Since visuals, such as charts, are inherently part of data stories~\cite{lee2015more}, it is essential to style them to make the data story more effective and appealing.
Typical actions include adjusting the layout of visual elements and applying animations.}}

In the final stage, the interviewees \textbf{communicate} their data stories with others.
After finishing and checking the story, the authors will \textit{share the story} informally or formally.
When sharing the story informally with the team, the authors often want to seek advice on improving the story quality, such as the core message's clearness and the content's completeness.
Then the authors may perform the previous tasks again to refine their story.
When the stories are shared formally, multiple actions can be taken according to the sharing format.
For example, if the sharing is a live talk or video recording, it is possibly necessary to write scripts and rehearse the presentation.
When the story is shared as a document, the authors may need to upload it to the cloud or send it to others through communication software.

\re{Though we attempt to divide the workflow into stages and tasks according to their chronological order, the procedure of data storytelling is not always linear.
Most of the time, the interviewees might experience conducting several tasks back and forth to polish their data stories.
For example, it is quite common that in the styling task, the authors could go back to prepare story pieces according to a more unified style.
Another example is that P2 often communicated with team members to review his stories in different stages and improved stories:}
\begin{displayquote}
\re{\textit{``The first thing is to really decide what you want to focus on, right?... 
Or which data you want to focus on showing... 
I actually use paper to take some notes or some ideas. And then I prefer to
discuss with someone else.''}}

\re{\textit{``and trying to make [findings] visual...
But then you still need to see
if they work into the framework that you want to explain.
Sometimes, maybe there is some step missing or sometimes some of the visuals are not clear enough.
So I also tried to discuss with someone and got some feedback.''
}}
\end{displayquote}

\re{\rem{Some participants}, under certain scenarios, might not conduct some tasks, as shown in Fig.~\ref{fig:task-role}(a).
They skip some steps according to the nature of their stories, such as the target audience and the occasion of communication.
For example, P5 mentioned that she might not integrate story pieces into slides or reports when discussing findings with her supervisor.
She instead showed several charts and presented the messages directly.
The reason was that they discussed a familiar project in an informal meeting.}

\subsection{Why AI is desired in data storytelling?}\label{sec:why}
\rem{This section summarizes the participants' common reasons why AI is desired. 
Table~\ref{table:why} shows an overview.}
\re{We noted that the reasons mostly reflect the participants' general beliefs and experiences about existing AI technologies while some proposed ways to leverage AI's advantages might be for future AI tools.
The comments should not be taken as absolute facts.}

\begin{table*}[]
\centering
\scriptsize
\caption{\revise{This table summarizes the reasons why AI collaborators are desired by our participants from the perspective of AI's advantages, example quotes, and the potential benefits to data workers.}}
\begin{tabular}{@{}p{4cm}p{2.3cm}p{5cm}p{5cm}@{}}
\toprule
\textbf{Reason }                                                                            & \textbf{Sub-reason}                                                                                                                & \textbf{Example quotes}                                                                                                         & \textbf{Potential benefits}                                                                                                                                                                                                     \\ \midrule
Elimination of repetitive and auxiliary tasks in data storytelling & /                                                                                                          & 
 \re{\textit{``I finish my own work well and leave PPT [slides] to the professionals, like AI... Why do I need to spend much time styling PPT [slides]?''}}                                                        & Data workers can save more time on the core of their work.
\\\midrule
\multirow{2}{4.0cm}{Potentially higher capability than humans in some steps of story creation}       & Broader commonsense and design knowledge coverage                                                                         &   \re{AI might explain \textit{``the reasons why the sales in one
year increase or decrease suddenly''} with additional knowledge.} & AI’s knowledge can compensate for data workers' lack of background information or design expertise for a more coherent and engaging story.                                                                                    \\\cmidrule{2-4}
& Accurate and fast computation                                     & \re{\textit{``AI might help us to find interesting combinations
between these ideas or it may be able to show us
some possible spaces which are not really explored
by our ideas...''}}                & Data workers can easily explore more possibilities of the stories with minimum effort.                                                                                                 \\\midrule
\multirow{2}{4.0cm}{Low time and financial cost of applying AI as collaborators for \re{data storytelling}}                  & Short turnover time & \re{Collaboration with AI is a \textit{``What You See Is What You Get''} experience.}      & AI can facilitate data workers’ job nature of quick iteration and reduce the risk of missing deadlines.    
\\\cmidrule{2-4}
 & Low expense &  
\re{\textit{``The cost [of humans] sounds higher than AI, including time and financial costs.''}}
& AI can be applied in daily work more frequently due to its low cost.
\\\midrule
Faithful and tireless AI collaborators in data storytelling.             & /                                                                                                       & \re{\textit{``If
[working with] AI, you can ask them whenever you
have new requirements.''}} & 
Data workers can propose new requirements to AI anytime without considering whether AI collaborators are willing and available.
\\ \bottomrule
\end{tabular}
\label{table:why}
\end{table*}

\subsubsection{AI reduces the workload of repetitive and auxiliary tasks}
A frequently mentioned reason for having AI collaborators is that they could reduce the workload of repetitive and auxiliary tasks.
The reason was mentioned by 14 participants (P1-P5, P7-P13, P16, P18).

Our interviewees often complained that they spent tremendous effort on repetitive and auxiliary tasks when creating appealing data stories.
P11 described such tasks as \textit{``dirty work.''}
Some example tasks included plotting charts, collecting data facts, and adjusting the layout.
P16 complained that she spent \textit{``most of the time''} adjusting the charts' or tables' styles according to her company's presentation template in the implementation stage.
She hoped that AI collaborators could
eliminate her efforts:
\begin{displayquote}
    \textit{``I finish my own work well and leave PPT [slides] to the professionals, like AI... Why do I need to spend much time styling PPT [slides]?''}
\end{displayquote}

Another example was mentioned by P7.
As a researcher, he sometimes gave talks at different lengths ranging from 5 minutes to 15 minutes depending on the occasion.
He needed several versions of slides then.
If AI automatically created slides according to the time requirement, P7 could avoid making slides several times.


\subsubsection{AI models are believed to have a higher level of capability than humans in some tasks}\label{sec:high_capability}
Among the 18 participants, 14 expressed that they would like to collaborate with AI since they believed AI might have a higher capability in specific tasks than humans.
They often mentioned two merits 
that they believed AI had and might potentially benefit data storytelling. 

First, some participants believed that AI could possibly \textit{have a broader knowledge coverage} over humans led by the large quantity of training data (P7-P10, P12-P14, P18).
P12 mentioned that humans had limited capability of searching and collecting information but AI could be more powerful.
Therefore, he expected that AI could give him more complete results when he asked them to collect data facts.
Similarly, P9 mentioned a case where AI might provide explanations for data facts leveraging additional knowledge outside the dataset, such as explaining \textit{``the reasons why the sales in one year increase or decrease suddenly.''}
P8 proposed that she could ask AI to generate presentation transcripts with different styles, such as humorous or formal ones.
\re{We considered P8's example as an instance of AI's broader knowledge coverage as writing humorous or formal presentation transcripts requires knowledge of writing styles.}

Secondly, the participants thought AI's higher computational power than human is likely to \textit{enable a fast and accurate search over the data or design space} (P1, P3, P4, P6, P10, P12, P14-P16, P18).
For example, P18 thought that AI could quickly create several data stories with different styles.
Then he could pick one style for sharing conveniently.
P6 mentioned the AI's potential in brainstorming ideas:
\begin{displayquote}
\textit{``AI might help us to find interesting combinations between these ideas or it may be able to show us some possible spaces which are not really explored by our ideas... AI should be really good at this kind of task.''}
\end{displayquote}

To utilize these advantages, another frequently proposed way was to ask them to provide inspiration to different tasks, such as selecting supportive data facts and choosing styles.
P1 further hoped that AI could help him check the problematic content, which might be led by data bias or incompleteness, or only provide a checklist.
His motivation was that humans could make mistakes since they were sometimes forgetful or lack of related knowledge.

\subsubsection{\re{Collaborating with AI may have less time and financial cost than with humans}}
When the participants talked about collaborating with AI and designers, an advantage of AI was the low time and financial cost.

First, they often appreciated that AI could have \textit{a quick response time}~(P7-P12, P15, P17), especially \textit{``when the schedule is tight''}~(P7).
P17 described the collaboration with AI as a \textit{``What You See Is What You Get''} experience.
P15 liked that AI could possibly allow him to iterate the data story swiftly.
In contrast, participants commented that working with designers or other human collaborators in data storytelling might take up to hours or days to finish tasks.
P9 further said that human collaborators could sometimes \textit{``miss their deadlines''}.
When working with AI, she did not have such concerns.

Furthermore, three participants (P8, P9, P11) thought AI could be cheaper than human collaborators' financial costs.
P11 said \textit{``The cost [of humans] sounds higher than AI, including time and financial costs.''}
Therefore, P8 believed AI might be more accessible to data workers than professional designers when handling less important tasks, such as preparing data stories for her regular team meetings.

\subsubsection{AI collaborators bear tasks faithfully and tirelessly}\label{sec:why_feelings}
According to our participants, it was common to polish the story pieces or styles several times until the authors felt satisfied with them.
Six participants thought AI was much easier to be asked to handle repetitive tasks or revise its outputs many times until data workers felt that the results totally matched their expectations (P4, P5, P7, P9, P14, P16).
The reason was that AI collaborators were more like \textit{``machines''} (P4).
Therefore, human users did not need to consider whether they were willing and available to handle the tasks.
For example, P4 commented that he could continuously ask AI to do some repetitive work without any worries.
When asking human collaborators to polish their work, the participants might feel reluctant or inconvenient since they needed to consider their collaborators' willingness and availability.
P16 compared the experience with human and AI collaborators:
\begin{displayquote}
    \textit{``You cannot continuously trouble others. If they did something incorrectly, you may ask them, `can you modify it here?' But when they continuously do something incorrectly, you may sometimes give up [asking them to revise] since you have already made several requirements or they become busy. If [working with] AI, you can ask them whenever you have new requirements.'' }
\end{displayquote}

Furthermore, the participants thought that AI collaborators could follow their ideas faithfully.
When collaborating with human collaborators, the participants might need to respect and compromise with others' preferences and opinions.
This point stood out when creating data stories since people tended to have their personal styles of storytelling.
P14 said:
\begin{displayquote}
    \textit{``When talking about styles, everyone may have different preferences. Therefore, I will respect [human collaborators' opinions] more. If I have some professionals, I believe that their designs can be appreciated more.''}
\end{displayquote}
When collaborating with AI, he could ask AI to make the story as close as possible to his intent.
P7 and P16 even thought that future AI could learn their personal preferences and create content as they did.

\subsection{Why AI is not a panacea for data storytelling?}\label{sec:why_not}


Throughout the interview, our participants provided various reasons for not collaborating with AI.
This section summarizes these reasons for why AI collaborators may not be favorable. 
\re{Again, these concerns reflect participants' general beliefs about AI and are heavily influenced by their experience with recent large-scale generative AI models. 
Some scenarios proposed by the participants might showcase their desire for future tools, such as the discussion between humans and AI.}

\begin{table*}[]
\centering
\scriptsize
\caption{\revise{This table summarizes the reasons why AI collaborators are not preferred by our participants from the perspective of AI's disadvantages, examples where AI is not preferred, and the potential concerns of data workers.}}
\begin{tabular}{{@{}p{2.5cm}p{2.5cm}p{6.5cm}p{5cm}@{}}}
\toprule
\textbf{Reason}                                                                             & \textbf{Sub-reason}                                                                                                                & \textbf{Example quotes }                                                                                                        & \textbf{Potential concerns} \\\midrule
Limited capability of creating data stories & / & \re{\textit{``Some AI-created figures may have counter-intuitive mistakes, such as drawing a person with one ear missing.''}} & The content created by AI collaborators cannot satisfy data workers' needs. \\\midrule
\multirow{2}{2.5cm}{Lack of understanding of data story context} & Missing project or data background & \re{\textit{``When you analyze a dataset, I feel somehow the important things are outside the dataset... I think AI may not capture [the background] well.''}}  &  Failing to capture the background may lead to a problematic goal or outline of the story. 
\\\cmidrule{2-4}
 & Ignorance of preferences & \re{\textit{``AI can generate the same output no matter who uses it''}} & The intent of data workers may not be expressed sufficiently expressed in the story.
\\\midrule
\multirow{2}{2.5cm}{Inconvenient to communicate with AI about data story designs and contents } & Requirement of precise instructions & \re{\textit{``It is hard for humans to specify the RGB values of a color red when [asking AI to] adjust colors.''}} &  It is challenging for humans to provide these precise values in their instructions to AI collaborators directly when authoring data stories. 
\\\cmidrule{2-4}  
& Limited interaction modalities &  \re{\textit{``However, the current AI
algorithms may only have one way of input. Then we possibly need to learn how to write prompts. Furthermore, we are hard to estimate whether AI understands what we want to do.''}} & Data workers cannot communicate with AI combining multiple familiar modalities conveniently and sufficiently. Miscommunication may happen.
\\\cmidrule{2-4}
& Limited ability of iterative discussion & \re{\textit{``[presenters] need to argue with others based on understanding the problem and the project requirements.''}} & AI may not replace data workers in discussing with the audiences. \\\midrule
Hard to enhance humans’ relationship with AI storytellers  & / &  \re{\textit{``It is more important
to let others feel that your team has its own insights,
a proper way to work, great characters, and good
communication skills in our pitches.''}} &
Having AI collaborators present data stories may not help build or even damage the relationship between data workers and clients.
\\\midrule
Considerable overhead of AI & / & \re{\textit{``If the AI model requires setting up the
deployment environment, it is quite annoying.''}} & The potential gain from collaboration with AI may not outweigh the cost.
\\\midrule
High reliance on training data & / & \re{\textit{``If you do not interview, there will be nothing in the database. How could you ask
AI to generate?''}} & Their lack of ability for collecting new data may affect the timeliness of data stories.
\\\midrule
Ethical concerns & / & \re{\textit{``AI cannot help me take the responsibility.''}} & There is a lack of regulations to deal with ethical issues.
\\\bottomrule
\end{tabular}
\label{table:why_not}
\end{table*}

\subsubsection{AI's capability of creation requires more improvements}\label{sec:why_not_capability}
13 participants had a deep concern regarding the performance of AI in creating data stories (P1-P3, P5-P8, P12-P17).
The potentially unsatisfactory performance has become an obstacle to introducing them to the production of data stories.
For example, P13 said:
\begin{displayquote}
\textit{``If [AI's] output can only achieve 40 to 50 marks [over 100 marks], I prefer to make drafts [of data stories] by myself or collaborate with other team members.''}
\end{displayquote}

Our participants mentioned their prior experience with AI's poor performance.
P15 distrusted AI-generated content since he had seen some misleading academic paper explanations created by ChatGPT.
He believed that such misleading content could be dangerous to users with limited expertise.
P8 expressed her opinion on AI's capability by saying, \textit{``some AI-created figures may have counter-intuitive mistakes, such as drawing a person with one ear missing.''}
P5 and P8 worried that AI could only produce a limited variety of voices when narrating data stories and the stories might not be engaging or interesting enough for their audience.

\subsubsection{AI may lack the understanding of data story contexts}\label{sec:why_not_context}
According to our participants, understanding the contexts of data stories was another challenging issue for AI.
Here contexts included both the \textit{project and data background} of the data story and the \textit{authors' and audiences' preference}.
It was a common viewpoint among our participants that AI was hard to take contexts into consideration.

Five participants considered that AI could not understand the project and data background (P1, P3, P6, P10, P12).
Among them, P1 mentioned the importance of the background of analytical projects and doubted whether AI could capture:
\begin{displayquote}
\textit{``When you analyze a dataset, I feel somehow the important things are outside the dataset. I feel the background is actually very important. When you only analyze the dataset about cars\footnote{The \textit{Car} dataset is a widely used example dataset. It can be found at \url{https://github.com/vega/vega-datasets/blob/main/data/cars.json}.}, [the findings] are very boring. However, under some specific scenarios, that
 dataset can be very important... I think AI may not capture [the background] well.''}
\end{displayquote}
P6 thought that AI might fail to understand the semantic meaning of domain-specific data, which was a challenging part of his consulting work.
He mentioned that he needed to meet his clients several times to understand the meanings of data with domain experts' help.

Regarding understanding the authors' and audiences' preferences, 
eight participants expressed their concerns (P3, P8, P9, P11, P12, P14, P16, P18).
P3, P12, and P14 did not think AI could understand their intent from their gathered data facts or story pieces.
P16 further discussed her difficulty in understanding her data story co-authors' intent when the feedback from them are unclear.
It was necessary to leverage her previous experience of working with them and discuss her understanding with the collaborators to align the ideas.
She considered that AI could not handle the case since AI required a \textit{``very clear idea''} as the input.
Another issue was that AI might have the same output regardless of the users, which might limit the expression of personal style (P9, P18).
P9 elaborated the idea by saying \textit{``AI can generate the same output no matter who uses it.''}


\subsubsection{AI models are hard to communicate as collaborators}\label{sec:why_not_communicate}
Our participants frequently mentioned that they found AI was not as easy to interact with as humans (P2, P8, P9, P11-P13, P15, P16, P18).
Therefore, they had doubts about AI collaborators in data storytelling.

The first issue is that the participants believe that AI needs \textit{precise instructions} when taking action (P9, P15, P18).
However, giving such instructions in data storytelling might not always be easy.
For example, P9 thought it was impossible to tell AI the concrete RGB values or the name of a color, so AI could hardly help her with the color adjustment task.
She said \textit{``It is hard for humans to specify the RGB values of a color red when [asking AI to] adjust colors.''}
P15 expressed a similar idea about general style adjustment.

A more critical issue is that most current AI models are limited to \textit{communicating with humans in specific modalities} (P8, P9).
For example, ChatGPT, a frequently mentioned AI system by interviewees, only communicates with humans using natural languages\footnote{More recently, advanced AI systems shed light on addressing the challenge of multi-modal communication between humans and AI, such as GPT-4o, with the ability to understand audio, image, and text input.}.
P8 considered that multi-modal communication was important in data storytelling:
\begin{displayquote}
    \textit{``When communicating with humans, many modalities can be applied. Despite using natural language to describe what you want, you can also show them pictures or your sketch. You can even use gestures to describe [your idea]. However, the current AI algorithms may only have one way of input. Then we possibly need to learn how to write prompts. Furthermore, we are hard to estimate whether AI understands what we want to do.''}
\end{displayquote}
She further pointed out that the lack of communication modalities might lead to miscommunication.
Even worse, humans could hardly realize the misunderstanding between humans and AI models.
Despite P8, P11 and P16 also expressed their concerns about miscommunication.

The last issue is that participants consider \textit{current AI to have limited capabilities to communicate with humans to reach a consensus when there is disagreement}
~(P8, P16).
Our participants thought this issue hindered AI's application in two-way communication.
P8 mentioned that AI could not have an insightful discussion with humans when \textit{``[presenters] need to argue with others based on understanding the problem and the project requirements.''}
She felt that AI lacked the ability to conduct \textit{``iterative''} discussions with humans and may not understand the goal of the conversation well. 
Such discussions could happen in pitch talks or project meetings.


\subsubsection{AI storytellers can hardly enhance humans' relationships}\label{sec:why_not_human_communication}
Another issue is that AI might not be able to help build relationships between storytellers and their audiences~(P11, P12).
According to P11, a business analyst,
data stories could play an important role in enhancing humans' relationships.
He mentioned that the communication of data stories between his team and clients helped to build trust:
\begin{displayquote}
\textit{``I think AI cannot replace the [communication] between humans and humans... It is more important to let others feel that your team has its own insights, a proper way to work, great characters, and good communication skills in our pitches.''}
\end{displayquote}
He indicated that the usage of AI in communication might not help create a good image of his team and therefore affect building partnerships with his clients. 
P12 further thought the application of AI might have a risk of impairing the relationship between him and his clients.
He considered that his data stories demonstrated his values as a professional to clients.
If his stories were made and communicated completely by AI, the clients might \textit{``question your value''} and have a lower willingness to work with his team.

\subsubsection{The overhead of applying AI is considerable}\label{sec:why_not_overhead}
Our participants pointed out that various overheads could hinder the wide application of AI collaborators
in data storytelling.

The first type of overhead is the \textit{learning cost of effective communication with AI}.
According to P13, the quality of AI's output depended on the quality of the communication between humans and AI models.
Therefore, it was critical to learn the proper ways of communication, which was echoed by P8's comment above.
P13 considered it necessary to have a \textit{``checklist''} to assist the interaction with AI.

The second type of overhead is \textit{configuring AI models}.
It might require a tremendous effort to configure AI models before involving them as collaborators.
P3, as an amateur user of generative AI models, complained about the overhead of deploying them,
\textit{``if the AI model requires setting up the deployment environment, it is quite annoying.''}
P12 described configuring AI's functionalities as \textit{``teaching''}  AI and said \textit{``if AI is not smart enough, I need to spend much time on teaching them. Then I prefer to do it myself or ask someone else to do it.''}
By saying it, P12 indicated that he needed to balance the overhead and benefit when collaborating with AI.
Similarly, P5 thought that it required some effort to configure AI to mimic her voice when asking AI to record a data story for her.
Therefore, she preferred to present the story herself.

The third type of overhead is led by \textit{AI's potential mistakes} (P3, P9).
Our participants mentioned humans needed to spend effort to correct AI's mistakes.
When the task itself was trivial, it might be even easier to conduct it by humans directly.
Therefore, the participants preferred to finish the simple tasks, such as adding annotations and moving the positions of visual elements, by themselves.

The unclear capability of AI models could also lead to the unnecessary \textit{overhead of trial-and-error}.
P15 and P16 indicated their concern that they might spend unnecessary time exploring 
whether AI could help them and received no assistance in the end.
Their time and effort in exploring AI's ability could become an overhead to the task. 

\subsubsection{The reliance on training data may limit AI's usage}\label{sec:why_not_data}
Our participants expressed their concerns about AI's reliance on training data.
They thought most of the existing AI models' performance often relied on large-scale training datasets built with public data.

First, general AI models might only be trained on public \textit{domain-agnostic data}.
As a result, seven participants worried that the general AI models could fail when dealing with domain problems (P3, P4, P6, P7, P11, P12, P16).
Our participants mentioned that some domain-specific data could hardly be collected, processed, and approached due to compliance issues.
Therefore, most of the commercialized AI systems might not access such data and lack the ability to handle it.

Second, AI cannot \textit{perform with the latest data} since their training data can be outdated and AI may not have the ability to collect new data.
P12 expressed his concern that AI trained on outdated data might be useless in his time-sensitive work.
P9 had a similar concern.
P17, as a journalist, pointed out:
\begin{displayquote}
\textit{``[The top journalists] need to conduct in-depth interviews. Each interview can last one hour. Each article can consist of ten to twenty interviews. Then they will organize the viewpoints and write an article with 10,000 words... If you do not interview, there will be nothing in the database. How could you ask AI to generate?''}
\end{displayquote}
She thought that AI was greatly limited by its ability to talk with humans and collect first-hand data to write compelling stories.
P7 also thought he had to collect data for AI.

Finally, since AI models can be trained with data collected from the web, \textit{the created content's reliability is questionable}.
P12 mentioned that most of the data on the web was not provided by the authority.
Since AI-created content could be based on such data,
it was hard to judge the content's accuracy and authority.
As a result, he could not use AI-created content.

\subsubsection{The application of AI can lead to ethical concerns}\label{sec:why_not_ethical}
When collaborating with AI, many participants thought that the potential ethical problems were not easy to be addressed.

Multiple participants (P1, P8, P10-P12, P14) raised their concerns about the \textit{responsibility} of the story content.
They might need to be responsible for AI collaborators' mistakes led by their insufficient capability or unreliable training data.
For example, P1 said \textit{``AI cannot help me take the responsibility.''}
If the responsibility issues were not well addressed, they would not feel confident to collaborate with AI.
P3 also worried about how to handle the \textit{copyright} of human-AI co-created content.

Another potential issue is data \textit{security} (P3, P12).
P12 thought that the data security problem hindered the broad usage of commercial AI systems.
He mentioned that his company limited the usage of AI due to the risk of data leakage.
He could only use the AI systems provided by his company even though their performance was unsatisfactory.

The final concern is the \textit{transparency} of AI models. 
P8 and P10 thought that they might have trouble knowing whether AI really understands their intent in making data stories.
P7 and P18 proposed that AI models should explain their rationales when humans felt skeptical about certain decisions.

\subsection{Where and how do humans want to collaborate with AI?}\label{sec:where_and_how}
\re{\rem{As shown in Fig.~\ref{fig:task-role}(b), the desired collaboration approaches depend on different tasks where humans would like to collaborate with AI.}
As a result, we report the results of the first two research questions together in this section with combinations of two independent dimensions, the \textit{tasks} of human-AI collaboration and the \textit{roles} of AI in collaborations.
Roles indicate how AI is expected to work with humans while tasks delineate what humans want to achieve through collaborating with AI.
This section introduces mostly users' expectations for future tools.
}


\subsubsection{What are AI collaborators' roles?}\label{sec:ai_roles}

According to our interviews, the desired roles of AI can be categorized into four types, \textit{creator}, \textit{optimizer}, \textit{assistant}, and \textit{reviewer}.
The characterization of roles reflects how humans would like to \re{distribute} the work in data storytelling between AI and themselves.
Examples of these roles are available in Table~\ref{tab:roles_tasks}

The first role of AI is the \textbf{creator}. 
When AI collaborators are assigned to a task in the workflow of data storytelling, they finish the entire task independently.
The responsibility of humans is limited to providing the raw materials as input and reviewing the AI-created output.
In other words, humans are not involved in the creation process.

The second role of AI collaborators is the \textbf{optimizer}. 
When AI collaborators act as optimizers, they do not handle the entire task but take the duty of fine-tuning humans' created content.
AI optimizers need to understand the input content and then improve the content.
In this collaboration mode, humans bear more workload and outsource optimization to AI.

The third role of AI collaborators is the \textbf{reviewer}.
Under this role, AI evaluates the task performance of humans and suggests potential issues or improvements.
Compared to AI optimizers, reviewers do not need to improve humans' created content directly. 
They only need to point out the problems or provide suggestions based on understanding humans' created content.
Then humans can decide how to fix the problem or which suggestion is adopted.

The fourth role of AI collaborators is the \textbf{assistant}.
Compared to the previous roles, AI collaborators take even less work when acting as assistants.
They are not authorized to modify anything created by humans and do not need to understand the content.
\revise{Instead, they provide assistance, such as recommending potential actions or answering users' questions, on specific sub-tasks according to humans' requirements or instructions.
The difference between assistants' recommendations and reviewers' suggestions is that reviewers provide suggestions based on the understanding of human-created content while assistants recommend potential actions based on humans' instructions.
Therefore, assistants do not need to understand human-created content.}

\begin{table*}[]
\centering
\scriptsize
\caption{This table provides \re{definitions and examples of AI's expected behavior under different roles} in all tasks of the data storytelling workflow. \re{Some combinations (indicated with italic fonts) of AI's roles and tasks were not proposed by our interviewees, which does not indicate that those combinations are not possible (see Sec.~\ref{sec:discussion_limitation})}}
\begin{tabular}{@{}p{3.2cm}p{3.2cm}p{3.2cm}p{3.2cm}p{3.2cm}@{}}
\toprule
 \textbf{Task}                         & \textbf{Creator}: the AI collaborator who finishes the task independently                                                                                                       & \textbf{Optimizer}: the AI collaborator who improves manually created content                                                       & \textbf{Assistant}: the AI collaborator who assists creation according to humans' requests                                                                                                                                             & \textbf{Reviewer}: the AI collaborator who evaluates manually created content and provides advice                                                                                                     \\ \midrule
Decide the core message   & Automatically decide the core message                                                                         & \re{\textit{No participant mentions AI’s expected behavior under the role in the task}}                                                & Summarize the goal from the analysis results, provide some potential ideas for inspiration                                                                & Summarize the resulting data story and check if the message in the summarization aligns with the author’s original idea \\\midrule
Collect data facts        & Automatically mine data facts from the dataset without instruction from humans                                  & \re{\textit{No participant mentions AI’s expected behavior under the role in the task}}                                               & Answer humans' questions about data or suggest data queries
& \re{\textit{No participant mentions AI’s expected behavior under the role in the task}}                                                                                                \\\midrule
Compile the story outline & Create an outline according to the current progress of the project                              & Improve the outline from the storytelling technique perspective & Recommend multiple outlines of telling the story for humans’ decisions                                                                                    & Check the logic flow of the outline                                                                           \\\midrule
Prepare story pieces      & Automatically plot charts or add text explanations                                                            & Polish the text or charts.                                      & Auto-complete code, support voice-based input                                                                 & Provide suggestions to improve the clarity, identify factual mistakes.                                      \\\midrule
Integrate story pieces    & Automatically generate the slides                                                                             & \re{\textit{No participant mentions AI’s expected behavior under the role in the task}}                                               & Manage the link between charts in slides and data                                                                                                         & \re{\textit{No participant mentions AI’s expected behavior under the role in the task}}                                                                                               \\\midrule
Style the data story    & Generate animated transitions, create decorative figures & Beautify the charts and tables        & Suggest layout, manage color usage                                                                                                                        & Evaluate the overall aesthetics and provide scores                                                            \\\midrule
Share the data story      & Create presentation videos                                                                                    & \re{\textit{No participant mentions AI’s expected behavior under the role in the task}}                                             & Upload the data story to a cloud server, distribute the data story through email                                                                              & \re{\textit{No participant mentions AI’s expected behavior under the role in the task}}                                                                                          \\ \bottomrule
\end{tabular}
\label{tab:roles_tasks}
\end{table*}

\subsubsection{What are AI collaborators' expected tasks?}\label{sec:ai_tasks}
\re{This section reports our participants' desired collaboration patterns, \ie, tasks and roles of AI, as summarized in Fig.~\ref{fig:task-role}(b).}

\re{\rem{Their desired collaboration concentrates} on the planning and implementation stages.
At the task level, the most frequently mentioned tasks include preparing story pieces (23 times), styling the story (23 times), and collecting data facts (12 times).
The number of times may exceed the number of participants since some participants mentioned multiple types of collaboration patterns in one task,~\eg, managing color usage as an assistant and improving aesthetics as an optimizer when styling story pieces (P3).
Notably, we would like to point out that the frequency counts reported in this paper only limit to our interviewees' opinions, instead of a panorama of all data workers. We hope future research can examine the results with a more representative group of data workers (see Sec.~\ref{sec:discussion_limitation}).}

\re{The distribution indicates that participants have different preferences for 
AI collaborators' roles at different stages.
The participants prefer AI collaborators as assistants in the planning stage to control the core idea and the outline.
Next, they would like AI to automate the implementation stage as creators or optimizers to offload their efforts while respecting their defined outline.
The rest of this section presents the most frequently mentioned tasks.}



\paragraph{Collect data facts}
Collecting supportive data facts is the most popular task for participants to collaborate with AI in the planning stage.
It was mentioned by 12 participants (P1-P4, P7, P9, P11-P13, P16-P18), where AI collaborators were expected to be assistants by 10 of them (P1-P4, P9, P11-P13, P16, P17).
Participants would like to ask AI assistants to collect and organize data facts from various sources.
Such work was often considered repetitive and time-consuming.
P12, as a business analyst, spoke about his expected AI assistant in collecting data facts from other documents:
\begin{displayquote}
    \textit{``When I make a report for a company now, I may need to search for [its annual reports] with search engines, let's say, annual reports of 2021, 2022, and 2020. After collecting these annual reports, I may find each year's income... Then I need to input each number manually to build the chart. This is a very cumbersome task... [The information] is very fragmented so you cannot get the numbers that you want at once. If AI can do something like collecting accurate income data in the past five years for me... I feel I can save lots of time collecting the data.''}
\end{displayquote}
The opinion was echoed by P8, who thought AI was good at summarizing existing information.

Besides collecting data facts, it was also desired to have AI for searching background information to enrich the data story. 
P1 expressed his need for an AI assistant to collect public opinions to illustrate the importance of his data story:
\begin{displayquote}
    \textit{``When I design the foreshadowing [for data stories], maybe it is possible to let [AI] write some stories. The stories are not fiction stories. I feel they are more like explaining why the topic [of a data story] is important... What [AI] does is more like collecting some public opinions or focuses.''}
\end{displayquote}
The help of AI could offload the efforts of collecting background information.

\paragraph{Prepare story pieces}
When preparing data story pieces, it was necessary to plot charts and write explanatory texts, which introduced a considerable burden to data workers.
Therefore, 15 participants mentioned that they preferred to receive help for this task.
The largest group of participants preferred to have AI create story pieces, such as writing texts or plotting charts (P5, P7, P8, P10-P12, P14-P16, P18).
They appreciated the collaboration where they input the outline and data facts, and AI could generate story pieces automatically.
Such collaboration was described with a metaphor by P14, \textit{``it is like that I prepare all ingredients for [AI] and ask them to cook dishes.''}
P16 further explained her opinion:
\begin{displayquote}
\textit{``I want to have AI to write texts... It can read the information in charts and analyze it. It is simple. For example, if the chart writes the average value of years or the yearly compound growth rate, then I don't need to calculate it, and AI can write [the related findings] itself based on the chart.''}
\end{displayquote}

Another two groups of users preferred to let AI optimize (P5, P13, P16-P18) or review their accomplished story pieces (P1, P12, P13) to improve the quality of story pieces or customize them according to the context of data stories.
For example, P13 sometimes felt unsure whether the story piece was hard to be understood by the audience. Therefore, he hoped that AI could provide suggestions:

\begin{displayquote}
\textit{``If we discuss the [research] problem with many lay people, their expected and data scientists' expected slide contents must be different. Under such scenarios, AI can provide more advice. For example, it can suggest that this chart is too complex, so you should break it into multiple charts to explain.''}
\end{displayquote}
P12 hoped that AI could review his story pieces using professional knowledge, such as statistical theories.

Some participants (P2, P3, P12, P13, P16) thought AI could assist story piece creation, such as 
auto-completing chart plotting code (P13),
and one-click translation (P16).
\re{P12 also envisioned that he could speak to AI rather than typing commands when creating story pieces.}

\paragraph{Style the data story}
Styling data stories often does not improve stories' content but enhances their appearance and facilitates understanding.
According to our interviews, common styling actions included adjusting the layout, changing color palettes, applying designated templates, and adding animations.
These actions were often cumbersome and could be non-trivial for data workers with a weak background in graphical design.
Therefore, aid from AI is welcomed.

Six participants preferred to have AI collaborators as assistants in styles (P1-P3, P6, P8, P9).
Potential assistance provided by AI included unified color palette management and suggesting layouts as the \textit{design idea} function in Microsoft PowerPoint~\cite{designidea} does.
P3 indicated the color usage in his domain was semantically meaningful since he often dealt with optical phenomena in chemistry experiments.
Therefore, he expected that an AI assistant could manage color palettes considering both semantic meanings and visual effectiveness.

Furthermore, seven participants expected that AI could create styles for their data stories (P7-P9, P11, P14, P16, P18).
Three participants (P1, P8, P17) implied that they frequently used the design idea function in Microsoft PowerPoint to suggest alternative styles for their slides.
However, its design suggestions were \textit{``limited''} (P17) and sometimes \textit{``outdated''} (P8).
As a result, P8 would like to further extend the function to transferring the styles in online well-designed slides to their own data story with AI collaborators as style creators.
P7 also mentioned a similar function.
Other examples of creating styles by AI included animating the presentation slides automatically (P9, P16).
Furthermore, P8 expected that generative AIs could create decorative or background figures for her data stories.

Lastly, AI might also optimize the existing style according to eight participants (P2, P3, P7, P8, P14-P17).
They hoped that AI could beautify the charts, tables, or the entire story.

\paragraph{Share the data story}
\re{In the communication stage, \rem{our participants preferred to collaborate with AI} for \textit{one-way} communication where storytellers might spread their data stories without direct interactions with their audiences.
Examples of one-way communication included spreading data videos or articles.}
Five participants expressed their willingness to ask AI collaborators to share the story \rem{on behalf of} them in one-way communication (P4, P7, P8, P10, P16).
For example, as a researcher on data storytelling strategies herself, P8 had an observation that \textit{``it is necessary to have multilingual videos when I want to spread my videos on worldwide platforms.''}
She thought AI could facilitate spreading data stories by automatically creating multilingual videos based on the original data story.
Similarly, P16 thought AI could create videos that would be delivered to the public \rem{on behalf of} her.
Furthermore, P10 and P16 preferred to have AI collaborators as assistants to upload their slides to the internal server or send them in an email to all attendees.
Therefore, they could save efforts from these auxiliary tasks. 

\section{What should future human-AI collaborative data storytelling tools look like?}\label{sec:suggestion}
\re{Our interviewees showed their passion for integrating AI into their workflow.
In this section, we recommend that AI collaborators in future data storytelling tools should be flexible and adaptive.
Tool developers should also consider the interaction between storytellers and audiences, provide natural interactions, and mitigate risks brought by AI.
For a comprehensive understanding of existing human-AI collaborative data storytelling tools, we refer our readers to a recent survey~\cite{li2024we}.}



\subsection{Flexible roles and tasks of AI collaborators in data storytelling workflow}

\rem{As Sec.~\ref{sec:workflow} indicates, 
unlike previous research on a specific task where humans and AI collaborate (\eg, the LookOut project~\cite{horvitz1999principles}), 
data storytelling is a multi-stage workflow consisting of various tasks.
The results of our interview suggest that data workers' expectations of AI collaborators' roles and tasks are diverse (see Fig.~\ref{fig:task-role}(b)).}

\re{We summarize two reasons for the diverse preferences.
First, the tasks in the data storytelling workflow often require different skills, while the skills of data workers are diverse.
As a result, data workers may rely on AI for unfamiliar tasks.
Second, the job nature may require users to have emphases on different tasks.
Users prefer to leave less important tasks to AI.}
\rem{Based on these findings, \textit{we suggest that tool designers should consider both users' skills and job nature when designing AI's roles in tools}.
The mapping from skills and job natures to AI's roles can be constructed through literature review and more empirical studies.}
\rem{Furthermore, \textit{future data storytelling tools should be modular and configurable to provide sufficient flexibility}.}
Data workers can decide where and how they would like to collaborate with AI systems instead of pre-defined workflows.
\rem{It requires the tool researchers and developers to define universal interfaces between modules and develop various modules that can communicate through these interfaces.
For example, the interface between the preparing story piece module and the integrating story piece module can be chart specifications.
If the preparation module is configured to be an AI creator, it will take data as input and return charts for the integration module.
When AI in the preparation module serves as an optimizer, it takes a chart created by humans as input and still send the improved chart to the integration module.
In this way, users can pick a suitable preparation module flexibly without affecting other modules in the tool.
It is also possible to equip AI creator modules for preparing story pieces with different AI techniques, such as LLMs to handle natural language input or rule-based approaches for deterministic results.}



\begin{figure}
    \centering
    \includegraphics[width=0.5\linewidth]{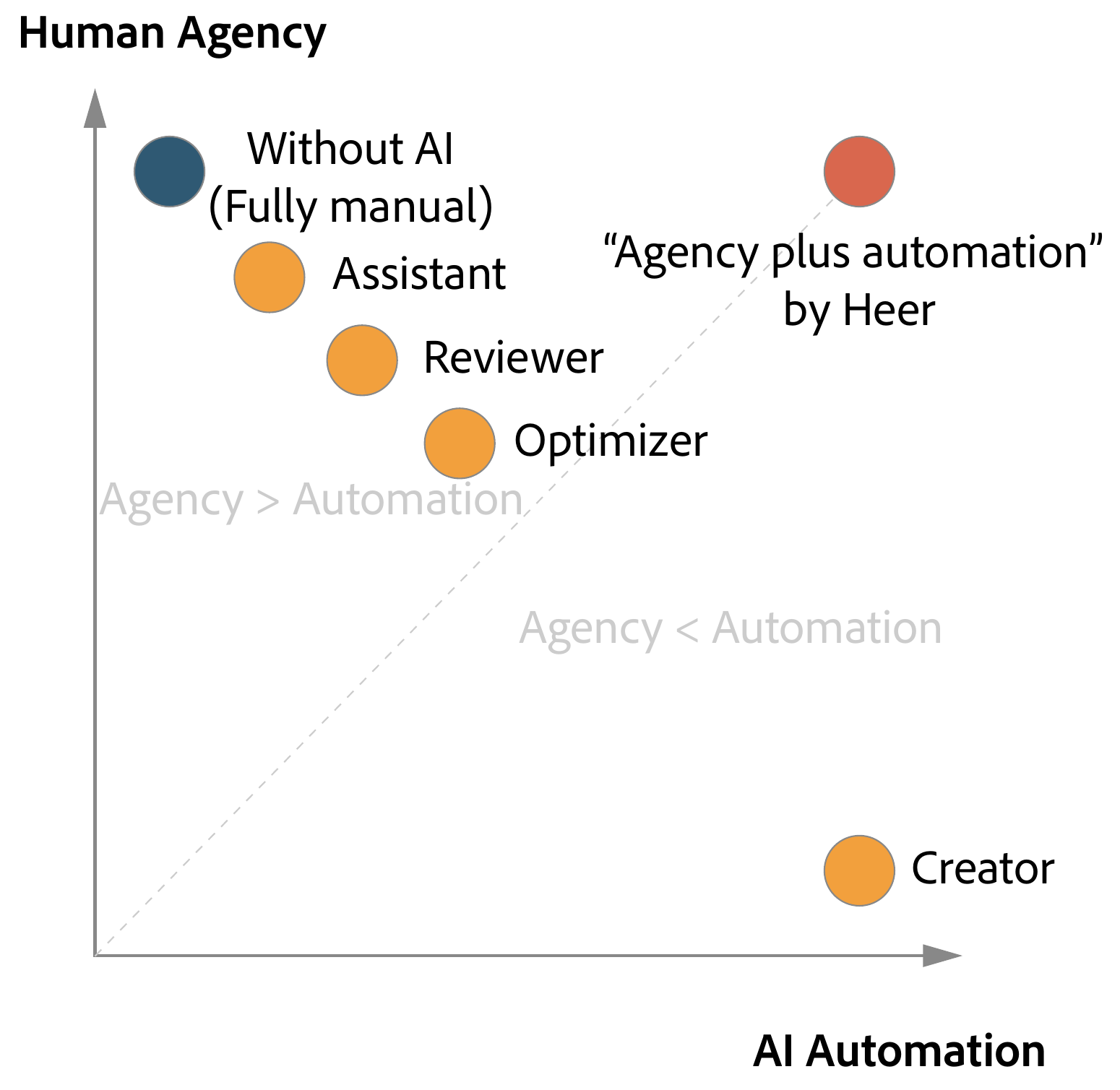}
    \caption{This figure demonstrates AI collaborators' roles against AI automation and human agency. We use fully manual story creation and Heer's ideal ``agency plus automation'' collaboration mode~\cite{heer2019agency} as two anchor points and derive the relative rough positions of other roles.}
    \vspace{-1em}
    \label{fig:role-automation-agency}
\end{figure}

Additionally, to help tool designers understand the roles and select suitable AI roles per users' tasks and requirements, we further assess the four roles using the \textit{``agency vs. automation''} framework~\cite{heer2019agency}, as Fig.~\ref{fig:role-automation-agency} shows.
\re{The framework is derived from the ideal goal of human-AI collaborative tools, achieving \textit{``agency plus automation''}, where both the automation brought by AI usage (\ie, \textit{automation}) and the control of humans (\ie, \textit{agency}) should be maximized as much as possible.
In this way, humans can spend minimum effort on tasks while maintaining a high level of control to achieve their target.}

\subsection{Aware of interactions between authors and audiences}
\rem{Data storytelling often requires interactions between data story authors (\ie, data workers in our research) and the audiences (Sec.~\ref{sec:why_not_human_communication}).
We propose that future AI-powered data storytelling tools should enhance \rem{the consideration of the interaction between authors and audiences with AI.} }


Inspired by Hancock~\etal~\cite{hancock2020ai}, we believe \textit{it is important to understand the authors' and audiences' perception of AI in their scenarios better before designing AI-powered data storytelling tools}.
For example, P11 and P12 implied that their audiences might not prefer AI's participation in their communication.
\rem{It is interesting to study which specific use cases of AI may affect audiences’ perceptions of data workers.
Based on the case studies, it is possible to derive concrete design guidelines for different industries or storytelling scenarios, such as team meetings or official presentations~\cite{brehmer2022jam}.
For example, according to our participants, when designing tools for users in financial industry, the communication stage should not involve AI.}
\rem{Furthermore, utilizing AI for two-way communication can be a potential direction.
Although AI collaborators might not be able to handle communication solely, they can assist humans for engaging presentations, \eg, the tools by Hall~\etal~\cite{hall2022augmented}.
For example, when a presenter is delivering the data story on stage, AI might be possible to answer audiences' various questions on behalf of the speaker, which extends the Ask Data feature in Tableau\footnote{\url{https://help.tableau.com/current/pro/desktop/en-us/ask_data.htm}} to be more interactive and flexible.
AI can explain a slide with more details or aggregate audiences' questions for the presenter's quick understanding and response.
}




\subsection{Adaptive to authors' and audiences' preferences}
\rem{As shown in Sec.~\ref{sec:why_not_context},
the lack of consideration for the preferences of authors and audiences may hinder the tools' acceptance. 
To address the concern, we propose that AI should be able to understand and follow various preferences.}

First, \rem{\textit{AI collaborators should understand the preferences of both authors and audiences using multiple sources, such as their past behavior and contextual information.}}
\re{According to Amershi~\etal~\cite{amershi2019guidelines}, users' preferences should be learned by AI implicitly.
The opinion is also echoed by P7 and P16 (see Sec.~\ref{sec:why_feelings}).
\rem{For example, AI can learn authors' preferred styles (\eg, color palettes, visual layout) and strategies for telling data stories (\eg, narrative structure) from their interactions and previous data stories.
The audiences’ preferences can also be inferred from their reactions to data stories, such as online comments to data videos or clapping hands for a story piece in an in-person meeting.}
Additionally, AI collaborator should be able to seek for information about preference automatically.
Due to data workers' lack of understanding of audiences~\cite{hou2017hacking,mao2019data,almahmoud2021teams}, they may not be able to capture the audiences' preferences comprehensively. }
\rem{For example, P7 imagined that AI could adjust the story based on the need for a formal or informal data story in different countries. To make suitable adjustments, AI collaborators need to collect more information about the context for data storytelling, such as culture-related knowledge. A potential solution is to automatically search for data stories under similar circumstances and apply their common designs.}

\re{Second, \textit{AI collaborators should help address the conflicts between authors and audiences}.}
It is unavoidable that the audiences' preferences and the authors' preferences may have conflicts. 
\rem{To handle conflicts, AI might present multiple versions of data stories to fit the different preferences of both parties at various levels.
AI can also act as potential audiences, allowing authors to experience the issues caused by conflicts.
In this way, human authors can decide on the final data story with a better understanding of the conflicts.}

\subsection{Cautious about misinformation and bias led by AI}
Different from fiction stories or creative writing, authoring data stories focus on communicating the data facts faithfully, completely, and effectively~\cite{diakopoulos2018ethics, mcbride2013new}.
For example, P15 mentioned that his charts for communicating data insights should be accurate rather than creative.
\re{As a result, we emphasize that the AI collaborator should avoid distorting data facts and introducing bias in data storytelling tools, as a reply to our participants' concerns, including the limited capability of creation (Sec.~\ref{sec:why_not_capability}), unreliable and potentially biased training data (Sec.~\ref{sec:why_not_data}), and other ethical issues (Sec.~\ref{sec:why_not_ethical}).}


To handle these concerns, \textit{data storytelling tool developers should be aware of the risks and design specific guardrails to eliminate potential misinformation and bias}.
Based on our \rem{participants' opinion}, we introduce several specific perspectives that should be taken care of.
First, AI collaborators should follow data workers' intent and purpose accurately and faithfully.
When there is any derivation from data workers' intent, data workers should be informed and allowed to make the final decision.
Second, AI collaborators should provide the source of data for human collaborators' awareness when utilizing external information and be able to remove all content without authoritative information sources per users' request.
Moreover, when AI is instructed to perform as creators, such as generating charts,
the result should be checked against widely applied visualization design and storytelling guidelines (\eg, \cite{bach2018narrative, chen2021vizlinter}) to avoid deceptive or misleading design or content.
For example, truncated vertical axes in charts should be checked and fixed.
\rem{One critical challenge in automatically checking data stories is how to represent diverse design guidelines in a way that machines can understand.}
\rem{Applying domain-specific language deserves more investigation~\cite{shi2024constraint}.}
Finally, as P1 suggested, AI collaborators could serve as reviewers to identify and correct human collaborators' mistakes.



\subsection{Effective and natural approaches to communicate and interact with AI based on data story characteristics}
The last suggestion in this section is about building convenient and natural communication approaches between data workers and AI in data storytelling tools.
\rem{As stated in Secs.~\ref{sec:why_not_communicate} and~\ref{sec:why_not_overhead}, inconvenient communication between humans and AI might lead to high learning cost and even miscommunication.
The problem is severe considering the multi-modal feature of data stories.
It is important to facilitate the communication needs for authoring different components in various modalities, such as charts, figures, animations, and infographics in the visual channel, as well as narration and background music in the auditory channel~\cite{shen2025reflecting}.}

\rem{First, 
\textit{tool designers should pay more attention to designing effective interactions considering the mutual influence between different components.}
For example, the specification of animation effects may be combined with narration authoring in natural languages to ensure their alignment~\cite{shen2024dataplaywright, wang2024wonderflow}.}

Second, \textit{the efficient combination of multi-modal interactions can be considered in future tools}.
\re{To make the communication between humans and AI closer to human-human communication, it is necessary to introduce multi-modal interactions.
Several research has revealed that multi-modal interactions can support convenient interactions between humans and data~\cite{srinivasan2020interweaving, walny2012understanding}.
We believe their experience can inspire the multi-modal interaction design in human-AI collaborative storytelling tools.
For example, users can sketch their preferred layouts and transitions, while expected emotions in the data story can be expressed with facial expressions or gestures.
Similarly, AI can present drafts to humans and ask questions.}

\section{Discussions and Limitations}\label{sec:discussion_limitation}
Our research is not without limitations.
\rem{Specifically, our methodology has limitations in sampling and analysis methods.
First, our results have limited coverage of data workers led by sampling bias.}
We noticed that most of our interviewees conducted research or white-collar jobs.
Furthermore, most of our participants are between 25 and 30 years old.
\rem{The imbalance between genders and the limited number of interviewees should also be aware.}
Therefore, \rem{the results reported in this paper may not cover opinions of all data workers and can be biased.}
Our participants' ideas should be treated as a probe into humans' attitudes toward human-AI collaborative data storytelling.
We hope our research can inspire future studies where other data storytellers, such as elders and YouTubers~\cite{yang2021design}, can be investigated.
\rem{Furthermore, during analyzing the interviewees' opinions, one author coded all transcripts and another reviewed the results.
Compared to analysis by multiple coders,
coding by one person can introduce more subjective bias.
To ensure that the bias is minimal, the author who was responsible for reviewing also conducted and attended several interviews to avoid being affected by the coder's personal bias.
Following prior practices~\cite{moore2023failurenotes}, the final codes were decided through iterative discussions between the two authors.}

Another limitation is the timeliness of our research.
Our interviews with data workers were conducted in the first quarter of 2023.
Along with the quick iteration of AI techniques, 
data workers' attitudes toward AI may change fast. 
Our participants' understanding of technological advances may also be obsolete.
Therefore, some of our results can be outdated in the future.
However, data workers' inherent needs (\eg, eliminating repetitive work) and concerns about AI (\eg, communicating with AI)  may not change.
As a result, we believe that our research is a starting point for raising researchers' awareness of data workers' attitudes toward human-AI collaboration, and has the potential to serve as a snapshot of social perceptions about AI-powered tools for reflecting the development of a fast-changing area in the future.
We sincerely hope that future research can continuously improve and complete our findings.

\re{We also realized limitations brought by the approach of interview study to identify potential AI usage in data storytelling.
Our summarized results reflect the expectation of AI participation instead of a complete view of potential AI usages in data storytelling.
As summarized in Table~\ref{tab:roles_tasks}, some combinations of AI roles and tasks were not mentioned by the participants.
They can also be explored in future tools.
For example, an AI optimizer can improve the chart and text organization in slide decks when integrating story pieces, while it can help polish presentation videos automatically, such as adjusting the pace of presentation.
When sharing data stories, AI reviewers can also provide suggestions to humans for improving their skills in presenting data stories.}

\revise{
Finally, our research focuses on summarizing data workers' desires for future human-AI collaborative tools instead of reflecting on actual experience with AI-powered storytelling tools.
We encourage participants to speak about their expectations and speculation about AI collaborators based on their practical workflow.
By doing so, we hope to ensure that they were not limited by the existing tools.
Another reason is that many existing data storytelling tools are research prototypes (\eg,~\cite{shi2020calliope}) and data workers may have limited experience.}

\rem{Based on the nature of an interview study and the purpose of investigating users' expectations, we acknowledge that our participants' opinions may include a mixture of assumptions and factual statements, influenced by their varying experiences and knowledge levels regarding fast-growing AI techniques. Therefore, their opinions should not be considered absolute facts but rather treated as references when designing future AI-powered tools.}
We would like to encourage field studies of how data workers use and perceive AI-powered tools or Wizard-of-Oz experiments, like McNutt~\etal~\cite{mcnutt2023design}, to collect their \rem{opinions on AI tools} based on hands-on experience.


\section{Conclusion and future directions}
\re{In this paper, we interviewed 18 participants from academia and industry about where, how, and why (not) they would like to collaborate with AI in their workflow.
The participants showed passion to having AI to perform four roles in the collaboration, including creator, assistant, optimizer, and reviewer for various tasks in the whole workflow.
The reasons of their willingness to collaborate with AI include the needs of reducing workload and AI's low time and financial cost.
However, they also pointed out rationales that hinders them from collaborating with AI in data storytelling, such as limited capability, lack of background understanding, and inconvenient communication between humans and AI.
Based on their opinions, we suggested that future data storytelling tools should be flexible and adaptive with convenient communication approaches between humans and AI.
They should also take AI's risks, such as misinformation, into consideration with a careful consideration of both storytellers and audiences. 
}

Our research can be extended from multiple perspectives.
First, it is worth more investigation about detailed preferences for human-AI collaboration, such as modalities of communication with AI.
\rem{Moreover, a fine-grained analysis of user preferences based on their AI and visualization literacy levels is promising, as their perception and evaluation of AI-powered tools can be directly affected by these levels.}
\re{Second, future research can dig deeper to understand the preferences for human-AI collaboration with more fine-grained workflows considering multiple dimensions, such as both task purposes and scenarios.}
Third, the roles of AI collaborators might be further studied and enriched 
based on our summarized four roles.
The interactions with different roles can also be explored for a deeper understanding of data workers' expectations.
\re{Finally, we noticed that most of our participants leverage slideshows for telling their data stories.
The investigation into practices of other data story formats beyond slideshows can provide a more comprehensive understanding of human-AI collaboration in data storytelling.}

\section*{Acknowledgments}
We would like to thank all interviewees for their insightful opinions and all reviewers for their constructive suggestions.
This work is partially supported by RGC GRF grant 16210722.

\bibliographystyle{IEEEtran}
\bibliography{main-abbr}

\begin{IEEEbiography}[{\includegraphics[width=1.0in,height=1.25in, clip,keepaspectratio]{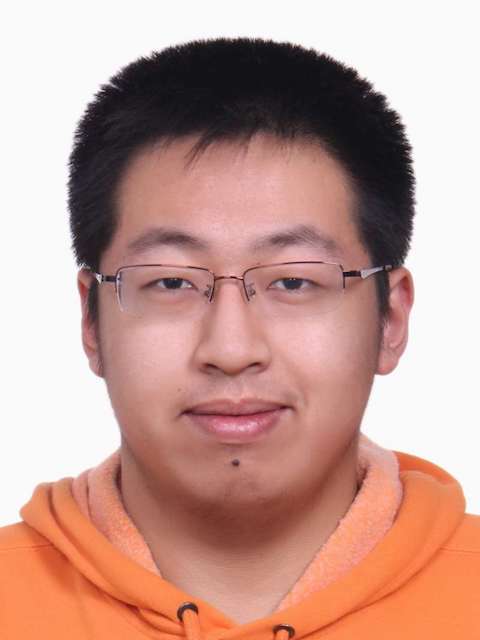}}]{Haotian Li} received his Ph.D. in Computer Science and Engineering in 2024 and his B.Eng. in Computer Engineering in 2019, both from the Hong Kong University of Science and Technology (HKUST). His primary research interests include data visualization and human-computer interaction. For more details, please refer to \url{https://haotian-li.com/}.
\end{IEEEbiography}

\begin{IEEEbiography}
[{\includegraphics[width=1in,height=1.25in, clip,keepaspectratio]{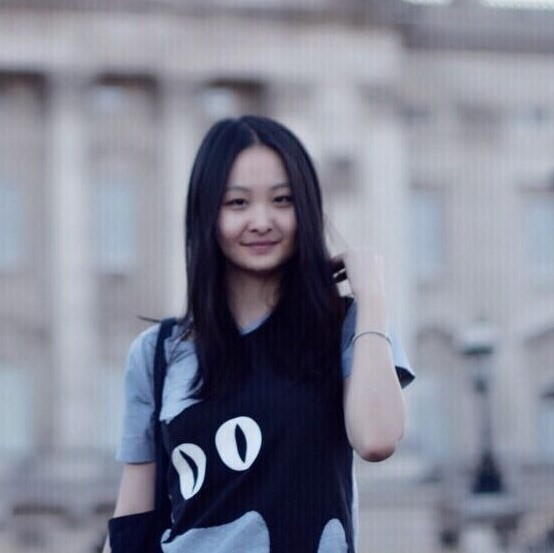}}]{Yun Wang} is a senior researcher at Microsoft Research Asia. Her research lies in the intersection of Human-Computer Interaction and Information Visualization. Her research focuses on enhancing human productivity and creativity through AI-powered systems for data exploration, information visualization, knowledge communication, and visual storytelling. For more details, please refer to \url{https://www.microsoft.com/en-us/research/people/wangyun/}.
\end{IEEEbiography}

\begin{IEEEbiography}[{\includegraphics[width=1.0in,height=1.25in, clip,keepaspectratio]{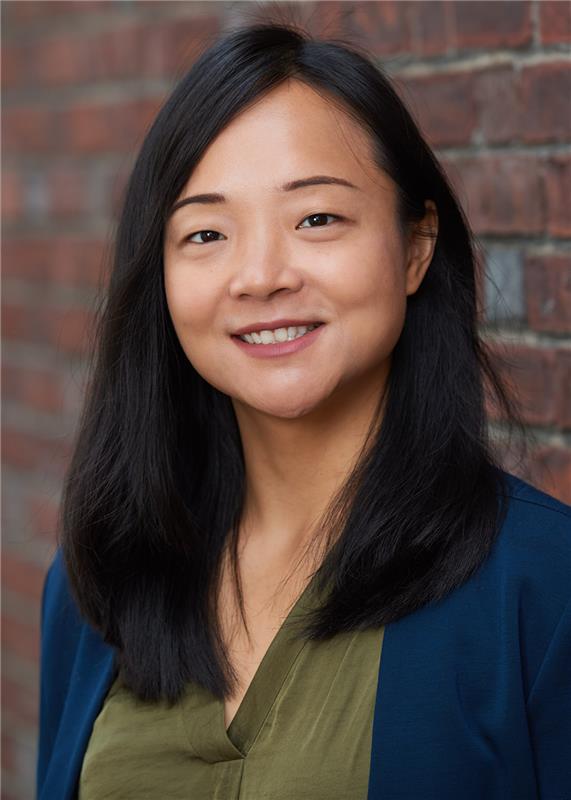}}]{Q. Vera Liao} is a Principal Researcher at Microsoft Research Montréal, where she is part of the FATE (Fairness, Accountability, Transparency, and Ethics in AI) group. Her current research interests are in human-AI interaction, explainable AI, and responsible AI, with an overarching goal of bridging emerging AI technologies and human-centered perspectives. Prior to joining MSR, she worked at IBM Research, and studied at the University of Illinois at Urbana-Champaign and Tsinghua University. She has authored more than 60 peer-reviewed research articles and received many paper awards at HCI and AI venues. She currently serves as the co-editor-in-chief for the Springer HCI Book Series, in the Editors team for ACM CSCW conference, and on the Editorial Board of ACM TiiS. She has also served as an Area/Subcommittee Chair and Senior PC member for CHI, FAccT, and IUI conferences.
\end{IEEEbiography}

\begin{IEEEbiography}[{\includegraphics[width=1.0in,height=1.25in, clip,keepaspectratio]{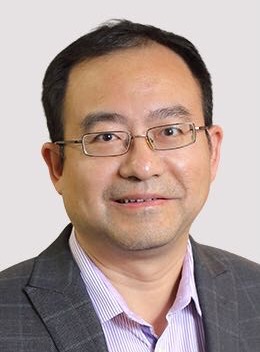}}]{Huamin Qu} is a chair professor in the Department of Computer Science and Engineering (CSE) at the Hong Kong University of Science and Technology (HKUST) and also the director of the interdisciplinary program office (IPO) of HKUST. He obtained a BS in Mathematics from Xi'an Jiaotong University, China, an MS and a PhD in Computer Science from the Stony Brook University. His main research interests are in visualization and human-computer interaction, with focuses on urban informatics, social network analysis, E-learning, text visualization, and explainable artificial intelligence (XAI). For more information, please visit \url{http://huamin.org/}.
\end{IEEEbiography}


\end{document}